\documentclass[11pt, reqno]{amsart}
\usepackage[margin=2cm]{geometry}                
\usepackage{verbatim}
\newcommand*\R{\mathcal{R}}
\usepackage{graphicx}

\usepackage{setspace}
\usepackage{units}
\usepackage{xcolor}

\usepackage{amsaddr}
\usepackage{hyperref}
\usepackage{xurl}
\usepackage{enumitem}

\title{The L\'evy Combination Test}
\author{Daniel J Wilson}
\address{Big Data Institute, Nuffield Department of Population Health,
University of Oxford, Li Ka Shing Centre for Health Information and Discovery, Old Road Campus, Oxford, OX3 7LF, United Kingdom}
\address{Department for Continuing Education, University of Oxford, 1 Wellington Square, Oxford, OX1 2JA, United Kingdom}

\begin{document}
\maketitle

\section*{Abstract}
\textbf{A novel class of methods for combining $p$-values to perform aggregate hypothesis tests has emerged that exploit the properties of heavy-tailed Stable distributions. These methods offer important practical advantages including robustness to dependence and better-than-Bonferroni scaleability, and they reveal theoretical connections between Bayesian and classical hypothesis tests. The harmonic mean $p$-value (HMP) procedure is based on the convergence of summed inverse $p$-values to the Landau distribution, while the Cauchy combination test (CCT) is based on the self-similarity of summed Cauchy-transformed $p$-values. The CCT has the advantage that it is analytic and exact. 
The HMP has the advantage that it emulates a model-averaged Bayes factor, is insensitive to $p$-values near 1, and offers multilevel testing via a closed testing procedure. Here I investigate whether other Stable combination tests can combine these benefits, and identify a new method, the L\'evy combination test (LCT). The LCT exploits the self-similarity of sums of L\'evy random variables transformed from $p$-values. Under arbitrary dependence, the LCT possesses better robustness than the CCT and HMP, with two-fold worst-case inflation at small significance thresholds. It controls the strong-sense familywise error rate through a multilevel test uniformly more powerful than Bonferroni. Simulations show that the LCT behaves like Simes' test in some respects, with power intermediate between the HMP and Bonferroni. 
The LCT represents an interesting and attractive addition to combined testing methods based on heavy-tailed distributions.
}

\section{Introduction}

A major problem for big data analysis is the penalty imposed on large-scale exploratory hypothesis testing by traditional methods such as the Bonferroni procedure. The main remedy for classical statistics has been the adoption of the false discovery rate (FDR) as an alternative to controlling the strong-sense family-wise error rate (ssFWER), regarded as gold standard control of multiple testing. Recently \cite{wilson2019harmonic} I employed Bayesian model-averaging arguments in the development of the harmonic mean $p$-value (HMP) procedure to show that testing \emph{groups} of hypotheses while controlling the classical ssFWER is an attractive alternative to controlling the FDR for \emph{individual} hypotheses. These two objectives appear to be closely connected, to the extent that they can be achieved simultaneously using a multilevel interpretation \cite{wilson2019tradeoffs} of Simes' test \cite{simes1986improved} (see also \cite{goeman2019simultaneous,goeman2021only}), which underlies the popular Benjamini-Hochberg FDR method \cite{benjamini1995controlling}.

Usually multiple testing is framed in terms of a number (say $L$) distinct pairs of null and alternative hypotheses. For example, there are $L$ genes and $L$ pairs of hypotheses concerning whether the observed expression level of each gene changes (or not) under two conditions. Yet often the same data are re-used to test different hypotheses. For example, there are $L$ genetic variants and $L$ hypotheses concerning whether a common set of observed outcomes are associated with each variant (or not). The latter scenario is connected to model choice and model averaging. The key idea implemented by the HMP and multilevel Simes procedure is to test both individual hypotheses and groups of hypotheses. Abstractly, the individual  null (respectively, alternative) hypothesis is that $p$-value $p_i$ follows the Uniform(0,1) distribution (or not), while the group null (respectively, alternative) hypothesis is that a set of $p$-values $\{p_i : i\in\R\}$ all follow the Uniform(0,1) distribution (or not). Rejecting a group of null hypotheses implies that one or more of the alternatives is true. One can then seek the smallest groups of hypotheses rejected at a pre-specified ssFWER \cite{wilson2019harmonic}.


The HMP exemplifies an emerging family of combined tests that exploits the properties of heavy-tailed Stable distributions. It is defined \cite{wilson2019harmonic,good1958significance} as
\begin{eqnarray}
\overset{\circ}{p}_\R &=& \frac{\sum_{i\in\R} w_i}{\sum_{i\in\R} w_i/p_i},
\end{eqnarray}
for some index set $\R$ of all the $p$-values $p_1 \dots p_L$ and weights $w_1 \dots w_L$. Remarkably, when small, $\overset{\circ}{p}_\R$ is approximately well-calibrated under the null hypothesis. Otherwise, for large $|\R|$, generalized central limit theorem shows that $\overset{\circ}{p}_\R^{-1}$ tends to the Landau distribution, a type of heavy-tailed Stable distribution \cite{wilson2019harmonic}. This enables computation of an `asymptotically exact' adjusted HMP:
\begin{eqnarray}
p^\star_{\overset{\circ}{p}_\R} &=& 1-F_{\textrm{Landau}}\left(\overset{\circ}{p}_\R^{-1}\middle | \log L+0.874, \frac{\pi}{2} \right) , \label{aeHMP}
\end{eqnarray}
where $F_{\textrm{Landau}}$ is the cumulative distribution function of the Landau distribution \cite{landau1944energy, landau1944translation}. Significance is assessed against a pre-specified threshold $\alpha$ at which level the ssFWER is controlled. 

The HMP procedure enjoys a range of desirable properties arising from the heavy tailed distribution of $1/p_i$. (i) It is robust to dependence, despite the derivation of Equation \ref{aeHMP} employing an independence assumption. Like the calibration property, robustness is better for smaller $\overset{\circ}{p}_\R$ \cite{wilson2019tradeoffs}. (ii) It controls the ssFWER while enabling arbitrary subsets of the $L$ $p$-values to be combined for the same pre-determined thresholds. (iii) The stringency of the significance threshold implied for the `raw' HMP $\overset{\circ}{p}_\R$ increases only logarithmically with the total number of tests $L$, compared to a linear penalty for the Bonferroni procedure. (iv) By construction it is interpretable as a likelihood-based analog to a model-averaged Bayes factor, where the optimal weights can be derived from parallel Bayesian and frequentist arguments \cite{wilson2019harmonic}.

The CCT \cite{liu2020cauchy} is an elegant alternative to the HMP. The CCT transforms $p$-values to Cauchy random variables using the transformation $\cot(\pi\,p_i)$ then sums the transformed $p$-values into a test statistic which is again Cauchy distributed under the null hypothesis:
\begin{eqnarray}
T_{\R} \,=\, \frac{\sum_{i\in\R} w_i \, F^{-1}_\textrm{Cauchy}\left( 1-p_i \right)}{\sum_{i\in\R} w_i}  \,=\, \frac{\sum_{i\in\R} w_i \, \cot\left( \pi\,p_i \right)}{\sum_{i\in\R} w_i} .
\end{eqnarray}
Unlike the HMP, the exact test is analytically tractable, producing combined $p$-value
\begin{eqnarray}
p_{T_{\{1:L\}}} \,=\, 1-F_{\textrm{Cauchy}}\left(T_{\{1:L\}} \right) \,=\, \pi^{-1} \cot^{-1}(T_{\{1:L\}}).
\end{eqnarray}
The stringency of the significance threshold for the test statistic $T_{\{1:L\}}$ is constant irrespective of the number of tests $L$, a potentially strong advantage. Like the HMP, the CCT builds in some robustness to dependence despite employing an independence assumption in its derivation \cite{liu2020cauchy}. For small values of ${\overset{\circ}{p}_{\{1:L\}}}$ and $p_{T_{\{1:L\}}}$, the HMP and CCT are approximately equivalent \cite{rustamov2020kernel,chen2020trade,fang2021heavy}, although there are important differences. Taken together, the HMP and CCT show that heavy-tailed Stable distributions have special advantages for combining hypothesis tests that are relevant to large-scale data analysis.

Despite their strong advantages, the CCT and HMP suffer from certain limitations explored in this paper. These limitations motivate the search for a test which combines their benefits and avoids their disadvantages. In seeking such a test, it is useful to frame the CCT and (in a limiting sense) the HMP as special cases of a general Stable combination test that employs the unique self-similarity or `fractal' property of sums of Stable distributions in which \cite{nolan2020univariate}
\begin{eqnarray}
X_i \,\overset{d}{=}\, X_\R \,=\, \frac{\sum_{i\in\R} w_i \, X_i}{W_{\R,\lambda}} + \Delta_\R ,
\end{eqnarray}
\begin{eqnarray}
W_{\R,\lambda} &=& \bigg[\sum_{i\in\R} w_i^\lambda\bigg]^{1/\lambda} 
\end{eqnarray}
and 
\begin{eqnarray}
\Delta_\R &=& \left\{\begin{array}{cc}
\delta\left(1 - \dfrac{W_{\R,1}}{W_{\R,\lambda}}\right) & \textrm{if $\lambda\neq1$} \\
\delta\left(1 - \dfrac{W_{\R,1}}{W_{\R,\lambda}}\right) + \frac{2}{\pi}\beta\gamma\left(\dfrac{\sum_{i\in\R} w_i\log w_i}{W_{\R,\lambda}} - \log W_{\R,\lambda} \right)
 & \textrm{if $\lambda=1$}
\end{array}\right.
\end{eqnarray}
In the above, $X_1 ,\dots X_L$ are assumed to be independent, identically distributed Stable random variables with tail index $0<\lambda\leq2$, skewness $-1\leq\beta\leq1$, scale $0<\gamma$, location $\delta$ and parameterization $\textrm{pm}\in\{0,1\}$ under Nolan's $S(\lambda,\beta,\gamma, \delta;\textrm{pm})$ notation \cite{nolan2020univariate}, with positive weights $w_1,\dots,w_L$. A Stable combination test is then devised (taking $\beta\geq0$) by defining $X_i = F^{-1}(1-p_i)$ for $i=1\dots L$ and 
\begin{eqnarray}
p_{X_{\{1:L\}}} &=& 1-F(X_{\{1:L\}}) . \label{SCTheadlinep}
\end{eqnarray}
In this notation, the CCT and (in the limit) the HMP correspond to Stable combination tests with standard Cauchy distribution $S(1,0,1,0;1)$ and Landau distribution $S(1,1,\pi/2,\log L+0.874;0)$ respectively. 

\section{The L\'evy combination test}
The L\'evy combination test (LCT) investigated in this paper is a Stable combination test arising from the standard L\'evy distribution $S(\frac{1}{2},1,1,0;1)$. The LCT compares the null hypothesis that a group of $p$-values all follow a $\textrm{Uniform}(0,1)$ distribution against the alternative hypothesis that one or more of them are enriched for $p$-values near zero. It is a multilevel test that controls the ssFWER while permitting all or any subsets of the $L$ $p$-values to be tested. The test statistic for the group of $p$-values indexed by set $\R$ is
\begin{eqnarray}
V_{\R} \,=\, \frac{\sum_{i\in\R} w_i \, F^{-1}_\textrm{L\'evy}\left( 1-p_i \right)}
{\left( \sum_{i=1}^L \sqrt{w_i} \right)^2}
  \,=\, \frac{\sum_{i\in\R} w_i \, \left[ \Phi^{-1}\left(\frac{1+p_i}{2}\right) \right]^{-2} }
{\left( \sum_{i=1}^L \sqrt{w_i} \right)^2} \label{defLCT}
\end{eqnarray}
and the combined $p$-value, adjusted for multiple testing, is
\begin{eqnarray}
p^\star_{V_{\R}} \,=\, 1-F_{\textrm{L\'evy}}\left(V_{\R} \right) \,=\, 2 \, \Phi\left(1/\sqrt {V_\R}\right) - 1, \label{defLCTp}
\end{eqnarray}
where $\Phi(x)$ is the standard Normal cumulative distribution function. The null hypothesis for index set $\R$ is therefore rejected when $p^\star_{V_{\R}} \leq \alpha$, where $\alpha$ is the target ssFWER.

The LCT enjoys a range of advantageous properties explored in the next sections. Like the CCT and unlike the HMP, the LCT is exact for any $L$ and has a convenient analytic formula. Like the HMP and unlike the CCT, the LCT avoids undesirable sensitivity to $p$-values near 1 and enables a multilevel test that controls the ssFWER. The LCT's multilevel test is uniformly more powerful than Bonferroni, overcoming a limitation of the HMP procedure. The penalty for expanding the total number of $p$-values scales more favourably than Bonferroni and less favourably than the CCT and HMP procedures. Although the exactness of the LCT depends on an independence assumption, it is more robust to arbitrary dependence than the CCT and HMP, possessing worst-case two-fold inflation for small significance thresholds when $L$ is large. However, the `headline' test that combines all $L$ $p$-values is less powerful.

In what follows, these properties are discussed in relation to the limitations of the CCT and HMP, revealing why the LCT occupies a unique position among Stable combination tests in combining the advantages of both, albeit at the loss of some power.

\subsection{Insensitivity to $p$-values near 1}
Despite its advantages over the HMP procedure in terms of its convenient formula and exactness for any number of constituent $p$-values, the CCT suffers the drawback of undesirable sensitivity to $p$-values at or near 1 \cite{chen2020trade,fang2021heavy,rustamov2020intrinsic}. Unfortunately this is probably a fatal flaw for the elegant CCT in many settings because generally $p$-values are defined conservatively such that $\Pr(p\leq \alpha | H_0) \leq \alpha$; they are said to be `superuniform' \cite{ramdas2019unified}. Consequently, $p$-values exactly equal to 1 are often legitimately encountered, for example with one-tailed tests, underpowered tests, $p$-values already adjusted for multiple testing and tests of discrete data.




One solution has proposed the use of a truncated Cauchy distribution to avoid problems with $p$-values near 1 \cite{fang2021heavy}. However, the resulting distribution of this test tends to the Landau distribution, making the proposed test asymptotically equivalent to the HMP procedure, which is more interpretable. In general, any Stable distribution test with both left and right heavy tails will be sensitive to $p$-values near 1, so an improved test must arise from the extremal Stable distributions ($|\beta|= 1$), which have only one heavy tail.

\subsection{Controlling the strong-sense familywise error rate} Another problem is that the CCT does not offer a straightforward closed testing procedure for controlling the ssFWER. The idea of the ssFWER is to control the family-wise false positive rate even in the presence of true positives. This is more difficult than controlling the false positive rate under the grand null hypothesis, and relevant in practical settings where one hopes to find true signals yet control the FWER.

Closed testing procedures (CTP) offer a general means to convert a well-calibrated combined test into a multilevel test that controls the ssFWER \cite{marcus1976closed}, but for practical use a shortcut procedure is required to avoid an infeasibly large number of tests. Consider a combined test based on Equation \ref{SCTheadlinep}, unadjusted for multiple testing:
\begin{eqnarray}
p_{V_\R} &=& 1-F\left( \frac{\sum_{i\in\R} w_i \, F^{-1}(1-p_i)}{W_{\R,\lambda}} + \Delta_{\R} \right)
\end{eqnarray}
where $p_{V_\R}$ controls the false positive rate in the sense that $p_{V_\R} \sim \textrm{Uniform}(0,1)$ if $p_i \sim \textrm{Uniform}(0,1)$ independently for all $i \in R$. A general CTP shortcut procedure can be devised by assuming the worst case scenario that $p_i = 1$ for all $i \in \R^\prime$. (Efficient methods for computing shortcut CTPs that improve power by avoiding this worst case assumption have been developed \cite{goeman2011multiple,dobriban2020fast,tian2021large} but the evidential basis for evaluating hypotheses in set $\R$ then depends on the quality of hypotheses outside set $\R$.) The adjusted $p$-value
\begin{eqnarray}
p^{\star}_{V_\R} &=& \max_{\mathcal{S}\subseteq \R^\prime}\left\{ 1-F\left( \frac{\sum_{i\in\R} w_i \, F^{-1}(1-p_i) + \sum_{i\in\mathcal{S}} w_i \, F^{-1}(0)}{W_{\R\cup\mathcal{S},\lambda}} + \Delta_{\R \cup \mathcal{S}} \right) \right\} 
\end{eqnarray}
controls the corresponding ssFWER in the sense that 
$\Pr(\cup_{\R\in\R_0} \, p^{\star}_{V_\R}\leq\alpha )\leq\alpha$,
where $\R_0$ is the index set of (usually unknown) true null hypotheses.

However, only Stable combination tests based on very-heavy tailed extremal Stable distributions (whose mean and variance are both undefined) can control the ssFWER through the shortcut CTP above because $F^{-1}(0)=-\infty$ except when $\lambda<1$ and $\beta=1$ \cite{nolan2020univariate}. This rules out the $Z$-test and CCT, which depend on the Normal distribution ($\lambda=2$, $\beta=0$) and Cauchy distribution ($\lambda=1$, $\beta=0$) respectively, but not the HMP procedure. Despite relying on the Landau distribution ($\lambda=1$, $\beta=1$), the inverse HMP converges to it only in the limit of large $L$, and is not an exact test. Therefore the HMP procedure has an advantage over an exact Landau combination test because it can control the ssFWER.

A family of very-heavy tailed extremal Stable combination tests that control the ssFWER is therefore obtainable by taking tail index $0<\lambda<1$, skewness $\beta=1$, scale $\gamma=1$ and location $\delta=0$ under parameterization pm=1, which implies that $F^{-1}(0)=0$, $\Delta_{\R}=0$ \cite{nolan2020univariate} and
\begin{eqnarray}
p^{\star}_{V_\R} &=& 1-F\left( \frac{\sum_{i\in\R} w_i \, F^{-1}(1-p_i)}{W_{\{1:L\},\lambda}} \right) . \label{padjSCT}
\end{eqnarray}
In this family, the L\'evy distribution ($\lambda=1/2$), named after the theoretical pioneer (e.g. \cite{levy1925calcul}), is the only Stable distribution with analytically tractable distribution function $F_{\textrm{L\'evy}}(x)=2[1-\Phi(1/\sqrt{x})]$, and this produces the LCT (Equation \ref{defLCTp}). The LCT thus shares the advantages of the CCT in terms of analytic tractability, ease of implementation and exactness (under independence) for any $L$, and shares the advantages of the HMP in controlling the ssFWER and insensitivity to $p$-values near 1.

\subsection{Uniformly better power than Bonferroni}
A limitation of the HMP procedure for controlling the ssFWER is that for testing small subsets of $p$-values, it is slightly more conservative than Bonferroni, despite much better power than Bonferroni when testing larger subsets \cite{wilson2019harmonic,wilson2019tradeoffs}. In contrast, the LCT is uniformly more powerful than Bonferroni for all subsets of $p$-values.

The Stable combination test of Equation \ref{padjSCT} is uniformly more powerful than Bonferroni for \emph{individual} hypotheses when 
\begin{eqnarray}
p^\star_{V_i} \,=\, \bar{F}\left( u_i^{1/\lambda} \, \bar{F}^{-1}(p_i) \right)
\,\leq\, p_i\,u_i^{-1} \,=\, p^\star_{B_i}, \label{pstarViLessThanpstarBi}
\end{eqnarray}
where $u_i = w_i^\lambda$ are Bonferroni weights that sum to one ($\sum_{i=1}^L u_i=1$) and $\bar{F}(x)=1-F(x)$. When $p_i > u_i$, this is true because $p^\star_{V_\R}\leq 1$ by definition. When $p_i\leq u_i$, rearrangement shows that Equation \ref{pstarViLessThanpstarBi} is equivalent to demonstrating $\bar{F}(x^{-1/\lambda})$ is concave in the range $x\geq0$ (assuming $0<\lambda<1$). Further, demonstrating that $p^\star_{V_i} \leq p^\star_{B_i}$ for all $i=1\dots L$ would be sufficient to demonstrate uniformly better power than Bonferroni because 
\begin{eqnarray}
p^\star_{V_\R} \,\leq\, \min_{i\in\R} p^\star_{V_i} \,\leq\, \min_{i\in\R} p^\star_{B_i} \,=\, p^\star_{B_\R}
\end{eqnarray}
for \emph{group} hypotheses indexed by $\R$.

The LCT can be shown to be uniformly more powerful than Bonferroni analytically because the second derivative of $\bar{F}(x^{-1/\lambda})$ is non-positive in the range $x\geq0$, demonstrating it is concave:
\begin{eqnarray}
\frac{\textrm{d}^2}{\textrm{d}\,x^2} \, \bar{F}\left(x^{-1/\lambda}\right) &=& \frac{\textrm{d}^2}{\textrm{d}\,x^2} \left( 2\Phi(x)-1 \right) \nonumber \\
&=& -2 \, x \, \phi(x) \quad\leq 0 \quad\textrm{when}\quad x\geq0 .
\end{eqnarray}
For other Stable combination tests, the second derivative of $\bar{F}(x^{-1/\lambda})$ can be evaluated numerically. This reveals that extremal Stable combination tests with $0<\lambda\leq0.5$ are uniformly more powerful than the Bonferroni procedure, but those with $0.5<\lambda<1$ are not (R code \href{https://doi.org/10.6084/m9.figshare.14535348}{doi:10.6084/m9.figshare.14535348}). Thus the LCT is the Stable combination test with the lightest tail index that incurs no loss of power relative to Bonferroni correction when testing individual or groups of hypotheses of any size.

\subsection{Robustness to dependence}
A major advantage of combination tests based on heavy-tailed Stable distributions is they exhibit robustness to dependence between the constituent $p$-values despite requiring no knowledge of the explicit dependence structure and despite relying on an independence assumption in their derivation \cite{wilson2019harmonic,wilson2019tradeoffs,liu2020cauchy,rustamov2020kernel,fang2021heavy,rustamov2020intrinsic,goeman2019comment,wilson2020generalized}. 



Superior robustness to dependence should be expected of the LCT compared to the HMP and CCT because of previous investigation of the generalized mean $p$-value (GMP),
\begin{eqnarray}
M_{r} &=& \left( \frac{p_1^r + \dots + p_L^r}{L} \right)^{1/r}.
\end{eqnarray}
Assuming independence between $p$-values, the limiting distribution of $M_{r}^{r}$ converges to an extremal Stable distribution with tail index $\lambda=-1/r$. GMPs with heavier tailed limiting distributions show better robustness to dependence \cite{wilson2020generalized,vovk2018combining}.

Specifically, $\frac{2}{\pi L} \, M_{-2}^{-2}$ converges to a standard L\'evy distribution $S(\frac{1}{2},1,0,1;1)$, the same distribution as the LCT statistic. This implies that the summands in the former behave like the summands in the latter, and indeed
\begin{eqnarray}
\frac{2}{\pi} \, p_i^{-2} &\approx& F^{-1}_{\textrm{L\'evy}}(1-p_i) \quad\textrm{as}\quad p_i \rightarrow0 
\end{eqnarray}
is a very close approximation. This relationship indicates that for large $L$ and small significance thresholds, the penalty on the number of tests increases with $L^{1/2}$ \cite{chen2020trade,wilson2020generalized,vovk2018combining}, and the Resnick-Davis condition for robustness to dependence applies \cite{davis1996limit}. Moreover, the L\'evy distribution-derived generalized central limit theorem threshold for $M_{-2}$ is two-fold inflated for small $\alpha$ compared to the threshold derived under worst-case dependence \cite{wilson2020generalized,vovk2018combining}, suggesting the same applies to the LCT.

Direct numerical analysis supports this claim. Following Vovk and Wang Theorem 8 \cite{vovk2018combining} and Embrechts and Puccetti \cite{embrechts2006bounds} enables computation of an upper bound on the inflation of the LCT under arbitrary dependence assuming large $L$ (R code \href{https://doi.org/10.6084/m9.figshare.14535339}{doi:10.6084/m9.figshare.14535339}). This showed that although inflation increased with $L$, for $\alpha \leq 0.05$, the worst-case inflation was very close to $2\,\alpha$, and increased slightly as $\alpha$ approached 0.5. For example, when $L=10^{10}$, worst-case inflation was $2.001\,\alpha$ at $\alpha=0.05$, rising to $2.132\,\alpha$ at $\alpha=0.469$.

\begin{figure*}[t]
\includegraphics[width=\textwidth]{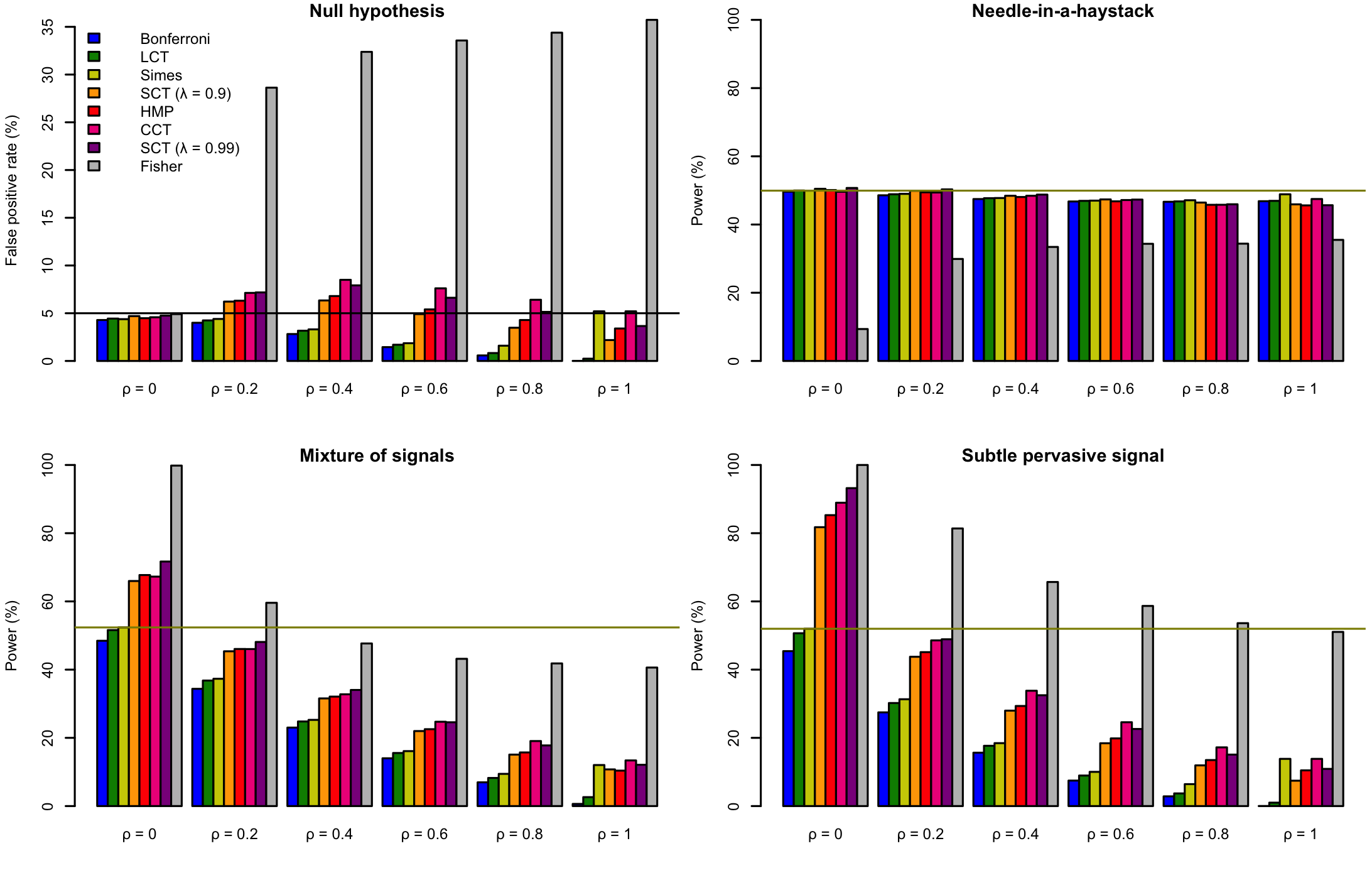}
\caption{\smaller{ `Headline' false positive rate (under the null hypothesis) and power (under `Needle-in-a-haystack', `Mixture of signals' and `Subtle pervasive signal' scenarios) when combining all $p$-values under a Wishart-Multivariate-Gamma model of dependence with varying strengths $0\leq\rho\leq1$. The tests were: Bonferroni (blue), LCT (green), Simes (yellow), SCT\textsubscript{0.9} (orange), HMP (red), SCT\textsubscript{0.99} (pink), CCT (purple) and Fisher (grey). Each bar represents 10,000 simulations. (R code \href{https://doi.org/10.6084/m9.figshare.14535351}{doi:10.6084/m9.figshare.14535351})
}}
\label{fig2}
\end{figure*}

Simulations under a Wishart-Multivariate-Gamma model of dependence relevant to likelihood ratio tests \cite{wilson2020generalized}, showed that in practice, robustness to dependence may be substantially better because  the LCT did not demonstrate any practically meaningful inflation (Figure \ref{fig2}). This contrasted with the CCT, HMP and two extremal Stable combination tests with $\lambda=0.9$ and $0.99$ (SCT\textsubscript{0.9}, SCT\textsubscript{0.99}), which showed some inflation at intermediate levels of dependence ($0.2 \leq \rho \leq 0.6$), and Fisher's procedure, which showed strong inflation except under independence ($\rho=0$). These simulations arguably provide a particularly stringent test of robustness to dependence because every constituent $p$-value was assumed correlated with every other. In fact, the LCT showed deflation under dependence, similar to that of the Simes test in the range $0.2\leq\rho\leq0.8$.

\subsection{Power of the LCT} Following \cite{wilson2020generalized}, I next compared the headline power of the tests to combine all $p$-values assuming $L=1000$ under three simulation scenarios: `Needle-in-a-haystack', `Mixture of signals' and `Subtle pervasive signal', in which 1, 100 and 1000 $p$-values were simulated under the alternative hypothesis with $Z$-statistics 3.0, 1.25 and 0.7 respectively (Figure \ref{fig2}). The remaining $p$-values were simulated under the null hypothesis.

The choice of simulation parameters led all methods to exhibit power close to 50\% under the Needle-in-a-haystack scenario except Fisher's method, which was substantially less powerful. However, the extreme inflation under dependence means Fisher's method would be impracticable except when independence could be safely assumed. Under the Mixture of signals and Subtle pervasive signal, the LCT exhibited similar power to Simes' test (50\% under independence, less under dependence) except under high dependence ($\rho\geq0.8$) when Simes' test performed better. All methods outperformed Bonferroni, but the SCT\textsubscript{0.9}, HMP, CCT and SCT\textsubscript{0.99} showed the best power (ignoring Fisher's method), and by an appreciably greater margin over the LCT than the LCT showed over Bonferroni. This illustrates the power-robustness trade-off inherent when comparing lighter versus heavier-tailed combination tests.

\begin{figure*}[t]
A \includegraphics[width=0.45\textwidth]{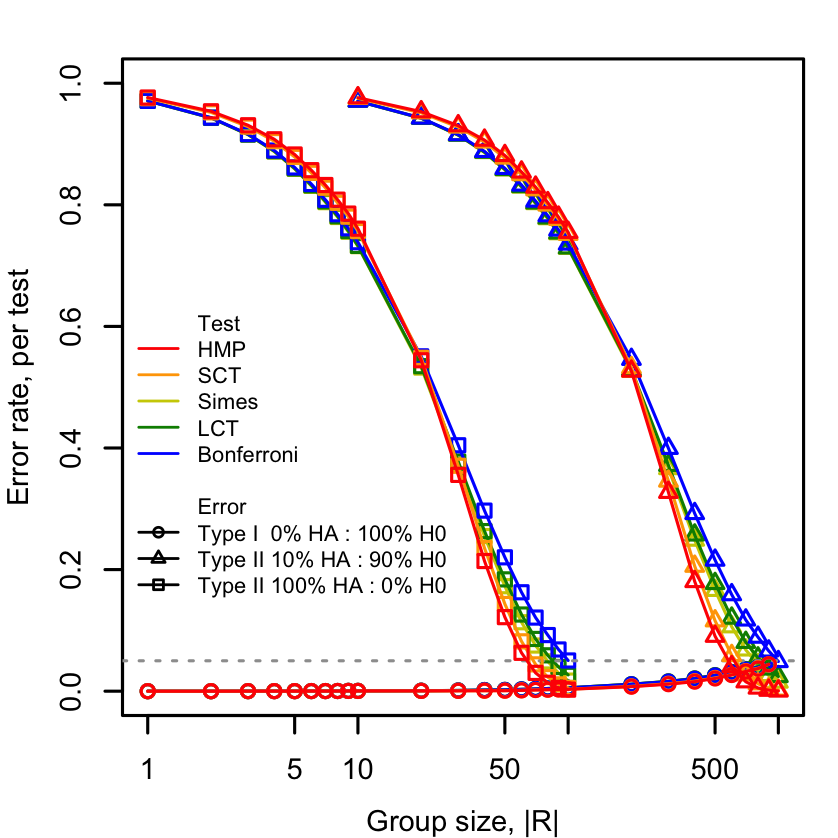} B \includegraphics[width=0.45\textwidth]{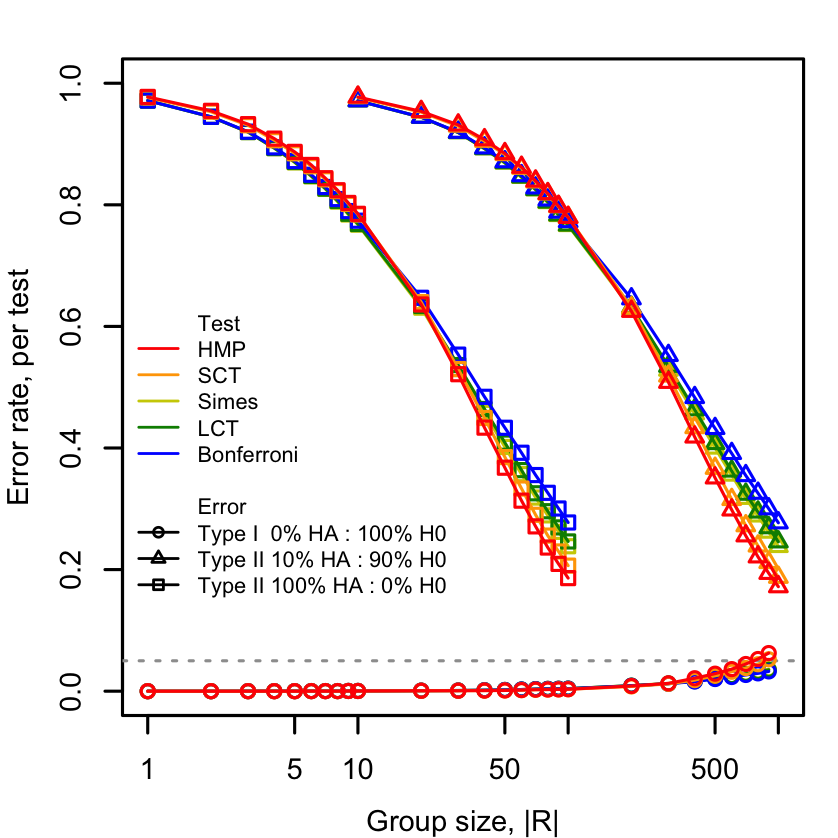}
\caption{\smaller{ Type I and type II error rates of multilevel tests under (A) independence (B) a multivariate normal model of dependence with $\rho=0.2$ \cite{goeman2019simultaneous}. Error rates were investigated in groups of different sizes ($|\R|$) and different mixtures of $p$-values simulated under the alternative (HA) and null hypothesis (H0) respectively. For each of 10,000 simulations under (A) independence and (B) dependence, I simulated a total of $L = 1,000$ normal random variables with means -2.0 and 0.0 under HA and H0, respectively, in proportion 100:900, calculating individual $p$-values by $Z$-test. The $p$-values were combined using the Bonferroni (blue), LCT (green), multilevel Simes (yellow) \cite{wilson2019tradeoffs}, SCT\textsubscript{0.9} (orange) and HMP (red) procedures. (R code \href{https://doi.org/10.6084/m9.figshare.14535354}{doi:10.6084/m9.figshare.14535354})
}}
\label{figMultilevelErrorRates}
\end{figure*}

For multilevel tests, different methods showed different performance characteristics for different sized subsets of the $L$ $p$-values (Figure \ref{figMultilevelErrorRates}). I investigated performance under independence and a multivariate normal model of dependence with $\rho=0.2$ previously shown to cause inflation for the HMP \cite{goeman2019simultaneous}. As reported previously \cite{wilson2019tradeoffs}, the HMP is substantially more powerful than Bonferroni and a multilevel Simes' procedure for large subsets of $p$-values ($|\R|/L>0.05$), but it incurs a slight reduction in power for small subsets of $p$-values ($|\R|/L<0.01$). This is not ideal, but tolerable because power is generally low for small subsets in any case. Extremal Stable combination tests with $0<\lambda\leq0.5$ do not incur this cost, unlike those with $0.5<\lambda<1$. For small subsets, SCT\textsubscript{0.9} showed a similar slight reduction in power to the HMP. In contrast, the LCT ($\lambda=1/2$) achieved power indistinguishable from Simes and Bonferroni. For large subsets, power increased as a function of the tail index $\lambda$, from Bonferroni (worst; $\lambda\downarrow0$), through the LCT ($\lambda=1/2$) and SCT\textsubscript{0.9}, to the HMP (best; $\lambda=1$). Simes test performed similarly to the LCT, indicating they occupy a similar performance trade-off between power and robustness to dependence. SCT\textsubscript{0.99} is not shown because it suffered drastic loss of power even for large subsets. Thus the HMP cannot be replaced by SCT\textsubscript{$\lambda\uparrow1$} for the purposes of multilevel testing and ssFWER control.









\section{Discussion}
In searching for a Stable combination test that combines the advantages of the HMP and CCT, the LCT appears to come closest, bringing analytic tractability and exactness under independence together with a multilevel procedure for controlling the ssFWER, insensitivity to $p$-values near 1 and interpretability in terms of model averaging. The LCT outperforms the HMP and CCT in terms of robustness to dependence, exhibiting close to twofold inflation under theoretical worst-case dependence, assuming the significance threshold is small and the number of $p$-values large. In simulations, there was no appreciable inflation. It also overcomes the limitation that the HMP is slightly less powerful than Bonferroni in the relatively underpowered scenario of testing small subsets of the $p$-values. However, the benefits of the LCT entail a cost in terms of reduced power compared to the HMP when combining large subsets of the $p$-values. If the imperfections of the HMP can be tolerated, it may therefore be preferred over all the other multilevel tests investigated here on the grounds of power. Further support for this position might be taken from the fact that the HMP provides the closest upper bound on a model-averaged Bayes factor of all the multilevel tests considered here \cite{wilson2020generalized}.

Stable distributions vary in their suitability for the important practical applications of combined testing. Distributions with left and right heavy tails, including the CCT \cite{fang2021heavy,rustamov2020intrinsic}, exhibit sensitivity to $p$-values near 1 which is a serious limitation in many practical scenarios. The class of extremal Stable distributions (skewness parameter $|\beta|=1$) that have just one heavy tail therefore appear more suitable. Among them, only the very heavy-tailed extremal Stable distributions (tail index $0<\lambda<1$), for which the mean and variance are both undefined, facilitate very convenient multilevel closed testing procedures for controlling the strong-sense familywise error rate. Within this class, only those for which $0<\lambda\leq1/2$ provide a convenient multilevel test uniformly more powerful than Bonferroni for all subsets of $p$-values. Since power generally increases with $\lambda$, the LCT ($\lambda=1/2$) therefore occupies a particular confluence of desirable properties. 

Interestingly, the inexactness of the HMP is a virtue that enables control of the strong-sense familywise error rate in a way not possible for the CCT nor a Landau combination test ($\lambda=1$). Indeed, multilevel procedures that control the ssFWER can be devised for GMPs for which $0<\lambda\leq2$ using generalized central limit theorem \cite{wilson2020generalized}. Like the HMP, which they generalize, these are not exact tests but assume large $L$. Aside from this important distinction, the investigation of Stable distributions here captures the limiting behaviour of a much larger class of potential combined tests based on sums of transformed $p$-values (see e.g.\ \cite{chen2020trade,loughin2004systematic,vesely2021permutation}). Indeed the LCT shares the limiting distribution of a transformed GMP\textsubscript{r=-2}. The inverse gamma distribution has been identified as another heavy-tailed distribution of potential interest \cite{fang2021heavy}. The standard L\'evy distribution is an inverse gamma distribution with shape parameter $1/2$, while sums of other inverse gamma distributions converge to other Stable distributions with tail indices defined by their corresponding shape parameters.

Coincidentally, the properties of the LCT appear to be similar to Simes' test in numerous aspects: both tests are analytically tractable and exact under independence, both possess empirical robustness to dependence, they offer multilevel tests to control the ssFWER for individual and group hypotheses, and both performed similarly in terms of power. Perhaps this is surprising given that Simes' test is order statistic-based while the LCT is sum-based, although the Simes test statistic does bound some GMPs \cite{wilson2019harmonic,chen2020trade,wilson2020generalized}. One notable difference is the Simes test's robustness to extreme positive dependence, which is shared with the CCT \cite{chen2020trade}. In summary, this investigation shows that the LCT offers an interesting and attractive test that reveals more about the family of heavy-tailed combined tests and, while the HMP may be ultimately be preferred in terms of power for practical purposes, the LCT and Simes test may be preferred in terms of robustness to dependence.



\section*{Acknowledgments}
\smaller{D.J.W. is funded by the Wellcome Trust (Grant 101237/Z/13/B) and the Robertson Foundation.}

\bibliographystyle{pnas-new}
\bibliography{bib}

\begin{thebibliography}{10}

\bibitem{wilson2019harmonic}
Wilson DJ (2019) The harmonic mean $p$-value for combining dependent tests.
\newblock {\em Proceedings of the National Academy of Sciences}
  116(4):1195--1200.

\bibitem{wilson2019tradeoffs}
Wilson DJ (2019) Trade-offs in model averaging using multilevel tests.
\newblock {\em Proceedings of the National Academy of Sciences}
  116(47):23384--23385.

\bibitem{simes1986improved}
Simes RJ (1986) An improved {Bonferroni} procedure for multiple tests of
  significance.
\newblock {\em Biometrika} 73(3):751--754.

\bibitem{goeman2019simultaneous}
Goeman JJ, Meijer RJ, Krebs TJ, Solari A (2019) Simultaneous control of all
  false discovery proportions in large-scale multiple hypothesis testing.
\newblock {\em Biometrika} 106(4):841--856.

\bibitem{goeman2021only}
Goeman JJ, Hemerik J, Solari A (2021) Only closed testing procedures are
  admissible for controlling false discovery proportions.
\newblock {\em The Annals of Statistics} 49(2):1218--1238.

\bibitem{benjamini1995controlling}
Benjamini Y, Hochberg Y (1995) Controlling the false discovery rate: a
  practical and powerful approach to multiple testing.
\newblock {\em Journal of the Royal Statistical Society. Series B}
  57(1):289--300.

\bibitem{good1958significance}
Good IJ (1958) Significance tests in parallel and in series.
\newblock {\em Journal of the American Statistical Association}
  53(284):799--813.

\bibitem{landau1944energy}
Landau LD (1944) On the energy loss of fast particles by ionization.
\newblock {\em Journal of Physics {U.S.S.R.}} 8(4):201--205.

\bibitem{landau1944translation}
Landau LD (1965) On the energy loss of fast particles by ionization in {\em
  Collected papers of {L. D. Landau}}, ed.{} ter Haar D.
\newblock (Pergamon Press, Oxford), pp. 417--424.

\bibitem{liu2020cauchy}
Liu Y, Xie J (2020) Cauchy combination test: a powerful test with analytic
  p-value calculation under arbitrary dependency structures.
\newblock {\em Journal of the American Statistical Association}
  115(529):393--402.

\bibitem{rustamov2020kernel}
Rustamov RM, Klosowski JT (2020) Kernel mean embedding based hypothesis tests
  for comparing spatial point patterns.
\newblock {\em Spatial Statistics} 38:100459.

\bibitem{chen2020trade}
Chen Y, Liu P, Tan KS, Wang R (2020) Trade-off between validity and efficiency
  of merging p-values under arbitrary dependence.
\newblock {\em arXiv preprint arXiv:2007.12366}.

\bibitem{fang2021heavy}
Fang Y, Tseng GC, Chang C (2021) Heavy-tailed distribution for combining
  dependent $ p $-values with asymptotic robustness.
\newblock {\em arXiv preprint arXiv:2103.12967}.

\bibitem{nolan2020univariate}
Nolan JP (2020) {\em Univariate Stable Distributions: Models for Heavy Tailed
  Data}.
\newblock (Springer).

\bibitem{rustamov2020intrinsic}
Rustamov RM, Majumdar S (2020) Intrinsic sliced {Wasserstein} distances for
  comparing collections of probability distributions on manifolds and graphs.
\newblock {\em arXiv preprint arXiv:2010.15285}.

\bibitem{ramdas2019unified}
Ramdas AK, Barber RF, Wainwright MJ, Jordan MI (2019) A unified treatment of
  multiple testing with prior knowledge using the p-filter.
\newblock {\em Annals of Statistics} 47(5):2790--2821.

\bibitem{marcus1976closed}
Marcus R, Eric P, Gabriel KR (1976) On closed testing procedures with special
  reference to ordered analysis of variance.
\newblock {\em Biometrika} 63(3):655--660.

\bibitem{goeman2011multiple}
Goeman JJ, Solari A (2011) Multiple testing for exploratory research.
\newblock {\em Statistical Science} 26(4):584--597.

\bibitem{dobriban2020fast}
Dobriban E (2020) Fast closed testing for exchangeable local tests.
\newblock {\em Biometrika} 107(3):761--768.

\bibitem{tian2021large}
Tian J, Chen X, Katsevich E, Goeman J, Ramdas A (2021) Large-scale simultaneous
  inference under dependence.
\newblock {\em arXiv preprint arXiv:2102.11253}.

\bibitem{levy1925calcul}
L{\'e}vy P (1925) {\em Calcul Des Probabilit{\'e}s}.
\newblock (Gauthier-Villars, Paris).

\bibitem{goeman2019comment}
Goeman JJ, Rosenblatt JD, Nichols TE (2019) The harmonic mean p-value: Strong
  versus weak control, and the assumption of independence.
\newblock {\em Proceedings of the National Academy of Sciences}
  116(47):23382--23383.

\bibitem{wilson2020generalized}
Wilson DJ (2020) Generalized mean p-values for combining dependent tests:
  comparison of generalized central limit theorem and robust risk analysis.
\newblock {\em Wellcome Open Research} 5:55.

\bibitem{vovk2018combining}
Vovk V, Wang R (2018) {\em Combining p-values via averaging}.
\newblock http://dx.doi.org/10.2139/ssrn.3166304.

\bibitem{davis1996limit}
Davis RA, Resnick SI (1996) Limit theory for bilinear processes with
  heavy-tailed noise.
\newblock {\em The Annals of Applied Probability} 6(4):1191--1210.

\bibitem{embrechts2006bounds}
Embrechts P, Puccetti G (2006) Bounds for functions of dependent risks.
\newblock {\em Finance and Stochastics} 10(3):341--352.

\bibitem{loughin2004systematic}
Loughin TM (2004) A systematic comparison of methods for combining p-values
  from independent tests.
\newblock {\em Computational Statistics \& Data Analysis} 47(3):467--485.

\bibitem{vesely2021permutation}
Vesely A, Finos L, Goeman JJ (2021) Permutation-based true discovery guarantee
  by sum tests.
\newblock {\em arXiv preprint arXiv:2102.11759}.

\end{thebibliography}

\end{document}